\documentclass{IOS-Book-Article}
\usepackage{float}
\usepackage{mathptmx}
\usepackage{caption}
\usepackage{graphicx}

\usepackage{soul}\setuldepth{article}
\usepackage[euler]{textgreek}
\def\hb{\hbox to 11.5 cm{}}
\usepackage{graphicx}
\begin{document}

\pagestyle{headings}
\def\thepage{}

\begin{frontmatter}              

\title{Long Short-Term Memory to predict 3D Amino acids Positions in GPCR Molecular Dynamics
}

\markboth{}{June 2021\hb}

\author[A]{\fnms{Juan Manuel } \snm{LÓPEZ-CORREA}%
\thanks{Corresponding Author: Juan Manuel LÓPEZ-CORREA, Univ. Politècnica de Catalunya. Barcelona Tech, 08034, Barcelona, Spain juan.manuel.lopez.correa@upc.edu.}},
\author[A,B]{\fnms{Caroline} \snm{ König}}
and
\author[A,B]{\fnms{Alfredo} \snm{Vellido}}
\hfill \break
\runningauthor{JM. LÓPEZ-CORREA et al.}

\address[A]{Computer Science Dept., Univ. Politècnica de Catalunya - UPC BarcelonaTech, 08034, Barcelona, Spain}

\address[B]{Intelligent Data Science and Artificial Intelligence (IDEAI-UPC) Research Center}

\begin{abstract}
G-Protein Coupled Receptors (GPCRs) are a big family of eukaryotic cell transmembrane proteins, responsible for numerous biological processes. From
a practical viewpoint around 34\% of the drugs approved by the US Food and Drug Administration target these receptors. They can be analyzed from their simulated molecular dynamics, including the prediction of their behavior in the presence of drugs.
In this paper, the capability of Long Short-Term Memory Networks (LSTMs) are evaluated to learn and predict the molecular dynamic trajectories of a receptor.
Several models were trained with the 3D position of the amino acids of the receptor considering different transformations on the position of the amino acid, such as their centers of mass, the geometric centers and the position of the $\alpha$--carbon for each amino acid. The error of the prediction of the position was evaluated by the mean average error (MAE) and root-mean-square deviation (RMSD). The LSTM models show a robust performance, with results comparable to the state-of-the-art in non-dynamic 3D predictions. The best MAE and RMSD values were found for the mass center of the amino acids with 0.078 Å and 0.156 Å respectively. This work shows the potential of LSTM to predict the molecular dynamics of GPRCs.
\end{abstract}

\end{frontmatter}
\markboth{June 2021\hb}{June 2021\hb}
\section{Introduction}
G-Protein Coupled Receptors (GPCRs) are a big family of eukaryotic cell transmembrane proteins. %
They are abundant cell surface receptors accounting for 4\% (800) of all human genes \cite{gpcrmd} and responsible for numerous biological processes.  This is the result of their ability to transmit extracellular signals, which makes them relevant for pharmacology and the research about drugs targeting these receptors. Around the 34 \% of the drugs approved by the US Food and Drug Administration  \cite{gpcrmd}. This has led, over the last decade, to active research in the field of proteomics.

The functionality of a protein depends widely on its 3-D structure, which
determines its ability for certain ligand binding. However, the 3-D structure of human GPCRs is not fully determined yet \cite{katritch2013structure}. 
As an alternative, when the information about the 3-D structure is not available, the investigation of the functionality of a protein can be achieved through the analysis of its amino acid sequence, which is known and available
in several public curated databases\cite{konig2013svm}.
Computing biotechnology, X-ray crystallography and cryo-electron microscopy over the last year has evolved exponentially, yielding 3-D models of many proteins. Such structures are publicly available at repositories, such as at the Protein Data Bank (PDB) \cite{databank}, but provide only a static view of the receptor's state. Molecular dynamics (MD) simulations are an interesting technology to explore dynamically the conformational landscape of receptors under the presence of different drugs. MD simulations take the 3D models as starting point for the computer assisted simulation and explore  the molecular dynamics at the simulated environment
\cite{latorraca2017gpcr}. Specific repositories of simulated MDs are available, such as the GPCRMD for MD simulations of GPCRs \cite{rodriguez2020gpcrmd}.

For the study of the temporal evolution of molecular dynamics with machine learning (ML) techniques, 
Recurrent Neural Networks (RNN) are used due to their capability for modeling temporal sequences. RNNs have shown success at applications such as human language modeling \cite{RCNN1}. In recent years, a particular RNN, Long Short-Term Memory (LSTM)\cite{LSTM4}, has been successfully applied to machine translation\cite{crossvalidation}\cite{LSTMreef1}, speech recognition \cite{LSTMreef2}, sequence learning  \cite{LSTMreef3} and weather forecasting \cite{weather}. LSTMs solve a limitations of the RNN architecture, the inability to  learn information originated from far past in time. LSTMs overcome that limitation by their ability to accumulate information for a long period of time and allowing the network to dynamically learn to forget old aspects of information. Recently, LSTMs have been used to mimic trajectories produced by simulations\cite {LSTM8}, achieving accurate predictions about a short time into the future.
LSTM and their variants have shown great potential in sequence processing \cite{LSTM5}, and there are several studies where they were applied for the analysis of trajectories from simulation systems \cite{LSTM6,LSTM7,LSTMmd}. Some authors, incorporate LSTM into the numerical integrator that solves Newton’s equations in molecular dynamics simulations \cite{LSTMnewton}. Another applies LSTM directly onto the low dimensional  molecular trajectories  and predicts the rare events in the sequential data \cite{LSTMmd,event}. However, none of them have reported robust results to predict the amino acid's position in the molecular dynamics of GPRCs.

In this work, the capability of LSTMs  are evaluated to learn and predict the 3D positions of the amino acids of a GPCR receptor in molecular dynamic simulations. In a first experiment, the prediction of two types of LSTMs are compared, the unidirectional and bidirectional variant of LSTMs. In the second experiment, different representations of the amino acid position and several variables of the LSTM are analyzed to find the combination of the parameters, which best predict the molecular trajectory. The experiments are carried out on a public available dataset of the molecular dynamic simulations of the  \textbeta2AR-rh1 GPCR receptor \cite{gcloud}.

The remaining part of the articles is structured as follows. In section Materials the dataset under study is explained. Methods section describe the ML model, data preprocessing and the experimental setup. The Result section evaluates the quality of prediction of the models per experiment. Finally in the Discussion and Conclusion section the results and impact of the study are discussed.

\section{Materials}

\subsection{MD simulations} 
In this work a dataset of the MD simulations of the \textbeta2AR-rh1 GPCR receptor is analyzed.
The \textbeta 2-adrenergic receptor (\textbeta2AR) is implicated in type-2 diabetes, obesity, and asthma, and is a member of the class A, rhodopsin-like GPCRs (rh1)  \cite{google}. 
This simulations have been created by \cite{gcloud} at the Google's Exacycle  cloud computing platform. The simulations under study in this work comprise 10.000 trajectories of the \textbeta2AR-rh1 GPCR receptor with a full agonist. Each trajectory describes the 3D position of the receptor during 28 consecutive timesteps, which are referred to as frames in this study. The time elapsed between each frame are 500 picoseconds. 
The receptor has 282 amino acids for which the position is predicted during the different frames of the simulation in this work.

\section{Methods}
\subsection{Long Short-Term Memory (LSTM)}

A specific and extremely popular instance of RNNs are LSTM \cite{LSTM4} neural networks, which show more flexibility and can be used for challenging tasks such as language modeling, machine translation, and weather forecasting \cite{RCNN4, ltranslationstm, LSTMreef1}. In this paper, unidirectional LSTM (ULSTM) and bidirectional LSTM (BLSTM) are used to predict the trajectories of the MD simulations. ULSTMs work by processing data in the forward direction, while BLSTMs  processes sequence data in both forward and backward directions with two separate hidden layers \cite{bidirect}. The bidirectional networks are often reported to yield better prediction results than unidirectional ones, such as at phoneme classification \cite{graves2005framewise}   or  speech   recognition \cite{graves2013hybrid}, to number a few. Bidirectional  LSTMs  have  not  been  used yet  in  molecular dynamic predictions of the GPCRs problem, based on a  review of the literature \cite{bi1, bidirect,bi2, bi3,bi4} . 

\subsection{Data Preprocessing}

 \ \  \  \  \  \  \ \textbf{\textit{ Data normalisation:}} The models are trained with normalized data. This process is done by applying a linear max–min normalization \cite{jahan2015state,asgharpour1998multiple}. The normalized data predicted by the model can by transformed back to the original range  of values to asses the quality of prediction in Angstrom (Å) units.

\textbf{\textit{Center of the amino acids:}}  Each simulation is made up of 28 frames (step positions) and 282 amino acids. 
However, the original information of the MD simulation provides the position of the atoms, not of the amino acids. For this reason, three representations for the 3D amino acid position are calculated: a) \textit{ Geometric Center } (CG), b) \textit{ Center of Mass} (CM) and c) \textit{ \textalpha--carbon} (\textalpha C) conforming three derived datasets with the 3D positions (xyz) for each amino acid in each step (frame) of the simulated molecular trajectory.
\break
\subsection{Experimental Setup}
In the following the variables and parameters used for the configuration of the experiments are explained: 

\textbf{\textit{nClones}}: As explained in the Materials section, the original dataset comprises 10.000 simulated trajectories of the  \textbeta2AR-rh1 receptor. Each of these trajectories is named a \textit{nClones} in this work.  Each trajectory is simulated under the presence of a full agonist.

\textbf{\textit{nSteps-in :}}  The training of the LSTM models is carried out with the information of short sequences of the 3D positions of the amino acids. The lengths of these sequences is named \textit{nSteps-in} in this work. In one step each amino acid has three position values (x,y,z). In the experiments the models are trained with different values for the parameter \textit{nSteps-in}.

\textbf{\textit{nSteps-out:}} After the training of the LSTM models with different\textit{ nSteps-in}, the model can predict a sequence of next steps for the trajectory. 
The number of predicted steps is referred to as the parameter \textit{nSteps-out}.


 \textbf{\textit{Model optimisation: }} Of the 10.000 \textit{nClones} available at the original dataset, 1.000 were taken for model creation. This dataset was split in 5 folds with 200 \textit{nClones} per fold. Four folds were used for training, conforming the train set and the remaining fold was used only for the evaluation of the quality of prediction of the model, conforming the test set. The model creation was carried out following a cross validation approach \cite{crossvalidation}. This means, the training was repeated 4 time per experiment. For each training repetition 3 folds of 4 train set folds were selected to train the model and the remaining fold (validation set) was use to test the predictions through \textit{Mean Average Error} (MAE) \cite{MAE}. This process was repeated for each fold of the train set. In this way, all the folds of the train set were tested without mixing the training and validation data. The training process generates 4 trained models and 4 testing results. The best model was chosen by the lowest MAE value. The best model was used to evaluate the predictive ability against new data never seen by the model,in this work the test set.

\textbf{\textit{Outline of experiments: }}
The first and second experiment were performed by two types of LSTMs -  ULSTM and BLSTM. The first experiment was carried out comparing the predictions of the LSTM on the CG, CM, \textalpha C transformed dataset. The experiments aims to find out which representation of the amino acid center is  best to predict the molecular dynamic sequences of the GPCR.

When a position sequence is shown to the LSTM, it needs to know the previous amino acid positions to predict the next positions. For this reason, the next experiment investigates if the parameter \textit{nSteps-in} has an impact on the quality of prediction. Values in the range of three, five and seven for the parameter \textit{nSteps-in} per center of the amino acid were evaluated. 

Finally, the third experiment was developed to know the capability of the ULSTM to predict different length of the sequence. For this experiments  the predictions of 12 \textit{nSteps-out} were evaluated. 

The results in the following section are reported using two metrics calculated on the original range of values in Angstrom (Å) unit:
\begin {enumerate}
    \item Mean absolute error (MAE) \cite{MAE} for each predicted value of the \textit{x},\textit{y} and \textit{z} position. 
     \item Root-mean-square deviation of amino acid positions (RMDS)\cite{RMSDme} for each predicted value of \textit{x},\textit{y} and \textit{z} position. 
\end{enumerate}

\section{Results}
In this section the results of the  experiments with LSTMs for the prediction of the sequences of the trajectories of the receptor are described. The  ULSTM and BLSTM were trained to predict multivariate time series data on sample trajectories. Table  \ref{table:centers ULSTM} and Table  \ref{table:BLSTM} show  the MAE (mean of the position "x", "y" and "z" for the amino acids) , MAEx (mean of the position "x"), MAEy (mean of the position "y"), MAEz (mean of the position "z") and RMSD by ULSTM and BLSTM respectively. In addition, in both tables the standard deviation \cite{std} (std) for MAE and RMSD is indicated.  In this experiment, to simplify the analysis the \textit{nStep-out} was set to a value of 1. About the \textit{nStep-in},  the results show the mean value from experiments carried out with the values 3,5,7  for the parameter \textit{nSteps-in}.

\begin{table}[h!]
\caption{ MAE predictions of the amino acid position by Geometric center, Mass center, \textalpha -carbon by ULSTM}
\centering
\begin{tabular}{ |c|c|c|c|c|c|} 
\hline
Amino acid center & MAE & MAEx & MAEy & MAEz & RMSD \\ [0.5ex] 
\hline\
Geometric &  0.0850 ± 0.024 &      0.0793 &      0.0864 &      0.0894 & 0.1703 ± 0.040\\
\textbf{Mass}      & \textbf{0.0781} ± 0.016 &      0.0728 &      0.0793 &      0.0822 & \textbf{0.1561} ± 0.027\\
\textalpha --carbon & 0.0792 ± 0.0219 &      0.0739 &      0.0809 &      0.0838 & 0.15934 ± 0.034 \\
\hline
\end{tabular}
\label{table:centers ULSTM}
\end{table}

\begin{figure}[H]
 
   \centering
   \includegraphics[width=10cm]{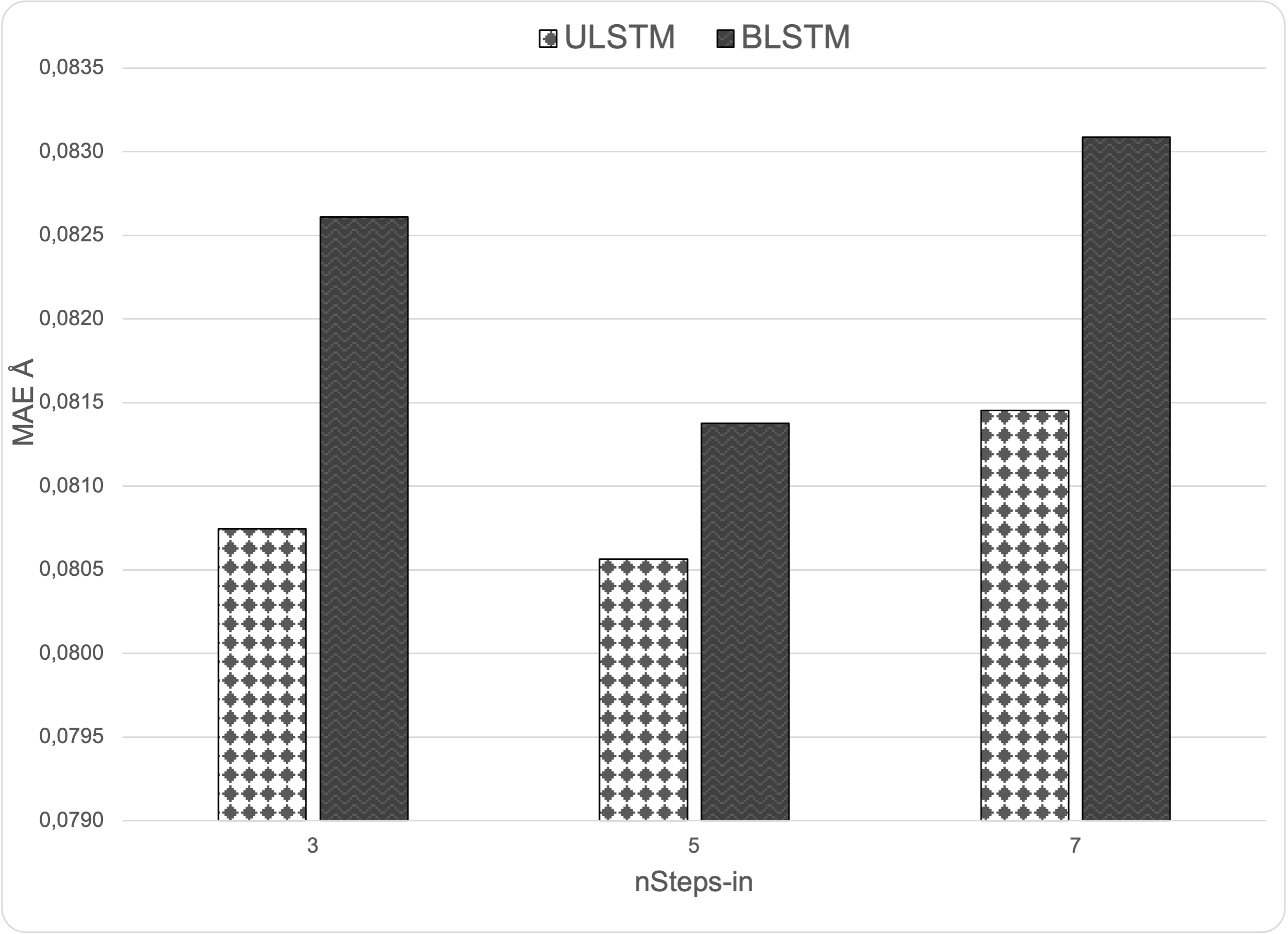}
  \caption{Mean Average Error(MAE) of the sequence prediction by ULSTM (full gray bar) and BLSTM (rhombuses bar) for three sequences lengths as\textit{nSteps-in} = 3, 5, 7 values. MAE in Å units }
  \label{fig:stepsin}
    \captionsetup{justification=centering,margin=2cm}
\end{figure}

\begin{table}[h!]
\caption{ MAE predictions of the amino acid position by Geometric center, Mass center, \textalpha -carbon by BLSTM}
\centering
\begin{tabular}{ |c|c|c|c|c|c|} 
\hline
Amino acid center & MAE & MAEx & MAEy & MAEz & RMSD \\ [0.5ex] 
\hline\
Geometric &     0.0835 ±        0.024 &      0.0775 &      0.0845 &      0.0883 & 0.1673 ± 0.040 \\
\textbf{Mass} & \textbf{ 0.0806} ±        0.021 &      0.0753 &      0.0821 &      0.0844 & \textbf{0.1615} ± 0.025 \\
\textalpha-carbon      &  0.0829 ±       0.023 &      0.0773 &      0.0842 &      0.0871 & 0.1661 ± 0.039 \\
\hline
\end{tabular}
\label{table:BLSTM}
\end{table}
Figure \ref{fig:stepsin} represents the results for the second experiment considering  the \textit{center of the mass} variable as amino acid representation.
The MAE of the sequence prediction are shown considering the \textit{nSteps-out} = 1 for  ULSTM (rhombuses bar) and BLSTM (full gray bar) with three sequences lengths as \textit{nSteps-in} = 3, 5, 7 values. 

The objective of the third experiment is the evaluation of the forecasting capability of a ULSTM to predict long sequences. The MAE and RMSD behavior for the 12 \textit{ nSteps-out} predictions is represented in the  Table \ref{table:length}. In this experiment the parameters, which have yielded the best results in the previous experiments were used, i.e. center of mass representation, \textit{nStepsIn} of 5 and a ULSTM. In this case, the MAE predictions is discriminated in 3D coordinates (MAEx, MAEy, MAEz). As well also, the standard deviation (st)) for mean 3D position MAE and RMSD is shown.

\begin{table}[h!]
\caption{ Error prediction by MAE and RMSD   for  12 length  sequences (\textit{ nSteps-out} ) with ULSTM. }
\centering
\begin{tabular}{ |c|c|c|c|c|c|} 
 \hline
 step &  MAE &  MAEx &  MAEy &  MAEz &     RMSD  \\ [0.5ex] 

\hline\
    1 &     0.0769 ±        0.0220 &      0.0713 &      0.0780 &      0.0814 & 0.1535 ± 0.0379 \\
    2 &     0.0814 ±        0.0224 &      0.0761 &      0.0823 &      0.0858 & 0.1626 ± 0.0381 \\
    3 &     0.0837 ±        0.0229 &      0.0784 &      0.0847 &      0.0880 & 0.1674 ± 0.0383 \\
    4 &     0.0842 ±        0.0231 &      0.0788 &      0.0855 &      0.0883 & 0.1684 ± 0.0388 \\
    5 &     0.0852 ±        0.0234 &      0.0797 &      0.0867 &      0.0892 & 0.1706 ± 0.0392 \\
    6 &     0.0854 ±        0.0234 &      0.0798 &      0.0869 &      0.0895 & 0.1712 ± 0.0390 \\
    7 &     0.0857 ±        0.0233 &      0.0797 &      0.0870 &      0.0903 & 0.1714 ± 0.0391 \\
    8 &     0.0862 ±        0.0236 &      0.0804 &      0.0870 &      0.0912 & 0.1725 ± 0.0396 \\
    9 &     0.0865 ±        0.0241 &      0.0810 &      0.0868 &      0.0919 & 0.1735 ± 0.0404\\
    10 &     0.0864 ±       0.0241 &      0.0802 &      0.0869 &      0.0920 & 0.1730 ± 0.0412 \\
    11&     0.08601 ±       0.02433 &     0.0793 &      0.0870 &      0.0916 & 0.1719 ± 0.0425 \\
    12 &     0.0861 ±      0.0246 &      0.0794 &      0.0872 &      0.0916 & 0.1722 ± 0.0433 \\

\hline
\end{tabular}
\label{table:length}
\end{table}

\section{Discussion}
The first experiment focusing on the different transformations of the amino acid positions has shown that center of mass is the transformation that best predicts the molecular dynamics sequences of the GPRC yielding minimum MAE and std values for ULSTM and BLSTM (bold numbers in Table \ref{table:centers ULSTM} and Table \ref{table:BLSTM}). However, the \textit{geometric center} and \textit{\textalpha -carbon center}, also, get quite good prediction results comparable with anothers works that work only with Static Molecular Structures \cite{nesterov20203dmolnet,hanson2009discovery}.
The second experiment seeks to discover the best sequence length as \textit{nSteps-in}. For both the ULSTM and the BLSTM  the minimum MAE was obtained for the value of 5 for the input steps. This means 5 steps as the best length for the input information of the sequence for the model to best predict future trajectories.  
Regarding the capabilities of ULSTMs and BLSTMs, the ULSTM demonstrate a better performance in all experiments.
Finally, the results of the third experiment reveal the increment of the error with the length of predicted sequence. With larger \textit{nSteps-out}  values, both the MAE and RMSD metric increase. In addition, the MAE in the plane "x" shows greater ability to predict more steps compared to the "y" and "z" plane. As well also, the plane z show the biggest
error analysing the 3D coordinates. 

\section{Conclusions}
GPCRs are family of receptors with great interest in pharmacology and molecular dynamics are a powerful tool to discover the conformational space and the behavior of the receptors. Due the large amount and complexity of data in MD simulations, machine learning approaches are a promising approach to discover relevant knowledge. This study has used a specific machine learning approach, namely LSTMs to study the ability to predict the movements of a receptor. This prediction is not trivial as the receptor comprises 282 amino acid, which yield a dataset of 846 data points. Furthermore, these datasets are in the context of a temporal sequence and methods taking into account the temporal evolution are needed. This study has demonstrated the potential of LSTMs to predict accurately molecular dynamics sequences of a GPRC receptor, specifically for \textbeta2AR-rh1. In addition, the  study has provided insights about which are the best parameters regarding the representation of amino acid positions, the lengths of the input sequence and length of the predicted sequence. In particular,  the \textit{center of the mass} is the best representation of the 3D amino acid  position for a complex receptor yielding the best results at the forecasting. Furthermore, the study has shown that the best length of input information are 5 steps. 
The prediction performance of ULSTM show slightly better results comparing with BLSTM, although both models achieved  accurate results. Finally, the study also confirmed that the capability to predict long sequences decreases with the lengths of the forecasted sequence. 
These results are important for the configuration of other experiments on the analysis of MD data. As a future line of research the use of generative models is planned in order to artificially generate MD trajectories. 
\section*{Acknowledgments}
This work is funded by Spanish PID2019-104551RB-I00 research project and by the PhD. training program (PRE2020-092428) through the Ministry Science and Innovation of Spain. 
\bibliographystyle{unsrt}
\bibliography{./samples.bib}

\end{document}